\documentclass[11pt]{article}
\usepackage{Blois}
\usepackage{epsfig}
\usepackage{graphics}
\usepackage{graphicx}

\def\be{\begin{equation}}
\def\ee{\end{equation}}
\def\bea{\begin{eqnarray}}
\def\eea{\end{eqnarray}}

\begin{document}
\vspace*{2cm}
\begin{center}
\Large{\textbf{XIth International Conference on\\ Elastic and Diffractive Scattering\\ Ch\^{a}teau de Blois, France, May 15 - 20, 2005}}
\end{center}

\vspace*{2cm}
\title{CONFERENCE SUMMARY}

\author{ Stanley J.~Brodsky$^{\dagger}$, Michael~Rijssenbeek$^{\ddagger}$ }

\address{$^{\dagger}$Stanford Linear Accelerator Center, Stanford University, Stanford, CA.\\ $^{\ddagger}$Department of Physics and Astronomy, Stony Brook University, NY.\\
}

\maketitle\abstracts{
Summary of the XIth International Conference on Elastic and Diffractive Scattering in Ch\^{a}teau de Blois, France, May 15 - 20, 2005, summarizing both theoretical and experimental presentations and discussions}

\section{Introduction}
Twenty years ago the first Blois Conference was organized by Basarab Nicolescu (University Paris VI) and Jean Tran Thanh Van (University Paris Sud) in the historic Ch\^{a}teau de Blois, France. The XIth conference in this international biannual series, which focuses on elastic scattering and diffraction, was organized again in Blois in May 15-21, 2005~\footnote{http://lpnhe-theorie.in2p3.fr/EDS05Accueil.html} by the original team plus Maurice Haguenauer (Ecole Polytechnique). 

Blois conferences have taken place all over the world: New York, Evanston, Isola d'Elba, Providence, Blois, Seoul, Protvino, Prague, and Helsinki, and the meeting has become a scientific forum for the community of researchers trying to unravel the foundations of elastic scattering and diffraction from first principles in quantum chromodynamics. Originally a very specialized field, it has now become one of the central topics in high energy physics, in particular because the measurements at HERA and Fermilab show that the number of diffractive events where a scattered proton remains intact in a high-energy inelastic collision constitutes a surprisingly large fraction of the entire rate: recent measurements from the ZEUS and H1 detectors at HERA show that approximately 10\% of the deep inelastic lepton-proton scattering cross section is diffractive (H.~Kowalski, M.~Soares).

\section{Elastic Scattering and Diffraction}
High energy elastic and diffractive scattering reactions are traditionally identified in Regge theory as due to Pomeron exchange in which the system exchanged between the projectile and target carries the quantum numbers of the vacuum. Events with large gaps in the rapidity distribution occur even in hard collisions involving very high momentum transfers; hard diffraction has now become firmly established, initially at the ISR and SPS at CERN in proton and antiproton reactions, and now most clearly at HERA in positron-proton collision and in proton-proton collisions at RHIC and the Tevatron. With the advent of QCD, hard diffraction is attributed to the exchange of two or more gluons with net zero color, and these processes have become an important observable for understanding fundamental aspects of the strong interactions. P.V.~Landshoff and A.~Donnachie reviewed the apparent dichotomy between the soft and hard aspects of pomeron exchange and the phenomenological manifestations, such as the remarkable growth with energy of the cross section for hard diffractive vector meson electroproduction. As discussed by C.~Ewerz, H-J.~Pirner, and D.~Schildknecht, theorists are now beginning to develop formalisms which encompass the transition between hadron and quark and gluon degrees of freedom and the duality between the two descriptions. QCD also predicts the existence of an odderon ($C$-odd) based on three-gluon exchange, which through interference with the pomeron exchange leads to remarkable charge asymmetries in diffractive reactions (K.~Itakaru, C.~Ewerz).

A particularly interesting nuclear diffractive phenomenon (D.~Ashery) is the demonstration of QCD `color transparency' by the E791 fixed-target experiment at Fermilab, which measured the diffractive dissociation of a high energy 500~GeV/c pion into two high transverse momentum jets while at the same time leaving the target nucleus remains intact. The Fermilab experiment confirmed the remarkable prediction, based on the gauge interactions of QCD, that the small transverse size  quark-antiquark Fock component of the pion projectile interacts coherently on every nucleon in the nucleus without absorption or energy loss, in dramatic contrast to traditional Glauber theory. The diffractive dijet experiment also provides crucial information on the quark-antiquark wavefunction of the pion. Other diffractive experiments which explore the structure of the photon are now in progress at HERA. 

A new understanding of diffractive events in deep inelastic lepton-proton scattering based on basic QCD mechanisms was discussed at the Blois meeting by G.~Ingelman and S.~Brodsky. Contrary to parton model expectations, the rescattering of the quarks with the spectator constituents shortly after it has been struck by the lepton has a critical effect on the final state in deep inelastic scattering as shown by P.~Hoyer et al. The rescattering of the struck quark from gluon exchange generates a dominantly imaginary diffractive hard pomeron exchange amplitude to the dijet. This produces a rapidity gap between the target and the diffractive system while leaving the target intact. The diffractive cross section measured in diffractive deep inelastic scattering can be interpreted in terms of the quark and gluon constituents of an effective pomeron as in the model by Gunnar Ingelman and Peter Schlein. Current parameterizations were presented by A.~Martin. Since the gluon exchange occurs after the interaction of the lepton current, the pomeron cannot be considered as a preexisting constituent of the target proton. The rescattering contributions to the DIS structure functions are not included in the target proton's wave functions computed in isolation and cannot be interpreted as parton probabilities. As discussed by G.~Ingelman, the resulting gluon exchange matches well to the phenomenology of the SCI (soft color interaction) model. As shown by D.~S.~Hwang, I.~A.~Schmidt, and Brodsky, gluon exchange in the final state also leads to the Bjorken-scaling Sivers single-spin asymmetry, a $T$-odd correlation between the spin of the target proton and the production plane of a produced hadron or quark jet. The connections between diffraction and coherent effects in nuclei such as shadowing and anti-shadowing are also now being understood (A.~Kaidalov, I.~Schmidt, S.~Brodsky). Diffractive deep inelastic scattering on a nucleon leads to nuclear shadowing at leading twist as a result of the destructive interference of multistep processes within the nucleus. In addition, multistep processes involving Reggeon exchange leads to anti-shadowing. In fact, as discussed by I.~Schmidt, since Reggeon couplings are flavor specific, anti-shadowing is predicted to be non-universal, depending on the type of current and even the polarization of the probes in nuclear DIS. 

\section{Saturation}
A central focus of the 2005 Blois conference was the physics of `saturation', a QCD phenomenon which limits particle production when the underlying gluonic scattering subprocesses significantly overlap in space and time (R.~Venugopalan, E.~Iancu; see also the work by St\'{e}phane Munier and Gregory Korchemsky). At very high energies, the gluon density is so high that two scatterings have the same probability as one. The theory of saturation is based on the Balitsky-Kovchegov equation and its extensions has analogs with stochastic methods used in other areas of statistical physics (E.~Iancu, R.~Enberg, G.~Soyez). The effects of saturation can be observed in the small x, high energy domain of deep inelastic lepton scattering at HERA, thus providing a window into nonlinear aspects of QCD. The theory of saturation predicts a parameterization of the HERA data  (`geometrical scaling'), which gives a remarkably good description of the deep inelastic structure functions at small $x$ in terms of a single scaling variable (R.~Venugopalan). The high occupation number of gluons can even lead to the formation of a `color glass condensate' (R.~Venugopalan, B.~Gay-Ducati), which may be causing a decrease of particle creation at forward rapidities in heavy ion collisions at RHIC, although the evidence is uncertain (M.~Strikman and A.~Capella.)

The nonlinear gluon interactions of QCD also underlie the physics of the hard (Balitsky-Fadin-Lipatov-Kuraev) BFKL hard pomeron which is postulated to control the energy dependence of hard reactions, as well as the distribution of particle production at very small value of $x$ and extreme values of rapidity. The extension of BFKL theory to high orders was discussed by L.~Lipatov, V.~Fadin, and J.~Bartels. An interesting analysis of the underlying energy transfer mechanisms based on laser backscattering was discussed by M.~Strikman.

Remarkably, Juan Maldacena's AdS/CFT duality between conformal gauge theory and string theory in 10 dimensions has begun to make an impact on QCD studies. The mapping of quark and gluon physics onto the fifth dimension of AntideSitter space is not only providing insight into the gluonium spectrum which controls the pomeron trajectory (Chung-I~Tan) and light-quark hadron spectroscopy (S.~Brodsky), but it is also able to explain the success of QCD counting rules for hard elastic scattering reactions using the methods of Joseph Pochinski and Matt Strassler.  As shown by Guy de~Teramond and Stan Brodsky,  the AdS/CFT  correspondence also provides an explanation and the dominance of the quark interchange mechanism in hard exclusive reaction, and it gives a model for the basic lightfront wavefunctions of hadrons which incorporates conformal scaling at short distance and color confinement at large distances. Lattice gauge theory is also making an impact on the QCD physics of high energy collisions as discussed by O.~Nachtmann, H-J.~Pirner, and E.~Meggiolaro.

\section{Experimental Work}
Much of the phenomenological work in Pomeron physics was pioneered by loyal participants of the Blois workshop series (J.~Bartels, A.~Donnachie, V.~Fadin, M.~Islam, V.~Khoze, P.V.~Landshoff, L.~Lipatov, A.D.~Martin, O.~Nachtmann, D.~Schildknecht, J.~Soffer, T.T.~Wu, and others). Many of the original participants of the first Blois Workshop also attended the 20th Anniversary Conference and presented their current work. While the discussions at the first Blois workshop centered on results from the CERN Intersecting Storage Rings (ISR) and the first data from the CERN $\bar{p}p$ collider (with predictions for the Tevatron collider and the now defunct SSC collider project), the XIth Blois Conference had speakers reporting the latest experimental results from the Fermilab Tevatron (K.~Goulianos, C.~Mesropian, C.~Royon), polarized $pp$ (and $p$C) using fixed and jet targets, and at RHIC at Brookhaven (S.~White, A.~Bravar, A.~Sandacz), and from HERA~II at DESY (M.~Kapishin, H.~Kowalski, M.~Soares, Y.~Yamazaki). HERA~II running has started and a significant and welcome statistical increase in diffraction data is expected before the machine closes in 2007.

The experimental efforts regarding forward Physics underway for the CERN LHC were extensively discussed (I.~Efthymiopoulos, K.~Eggert, G.~Ruggiero); the large increase in $x$ and $Q^2$ reach in $pp$ and AA collisions will provide significant tests and advances for QCD-inspired diffraction phenomenology and calculations. There was also speculation that the Froissart bound for the total $pp$ cross section will be saturated at the LHC, but with its value controlled not by pion exchange, but by by the exchange of the lightest glueball (H.~Meyer) as originally predicted by B.~Nicolescu. An exciting future prospect at the LHC is the possibility to observe the Higgs boson in doubly-tagged diffractive collisions $pp \rightarrow p + H + p$ at the LHC (K.~Eggert, A.D.~Martin, N.~Zotov, J.~Soffer); the Higgs would be found as peak in the missing mass spectrum, rather than in a specific decay channel. The novel possibility of black hole formation at the LHC was discussed by G.~Landsberg.

A number of presentations addressed the status of new and existing cosmic ray detection arrays (P.~Billoir, K.~Kasahara), where the discrepancy above the Greisen-Zatsepin-Kuzmin cutoff between the AGASSA excess and HiRes falloff continues to generate interest. Some proposals for very forward instrumentation at the LHC are dedicated to providing benchmarks for the cosmic ray shower initiation in simulations: Castor in Totem/CMS for forward electromagnetic showers, and LHC in ATLAS (Y.~Muraki) for the measurement of forward (zero-degree) $\pi^0$'s. Zero-degree neutron calorimeters, which operate successfully (and function also as heavy ion luminometers) at the RHIC experiments, will have their counterparts at LHC (R.~Schicker, S.~White); they will be used for the LHC heavy-ion program but may also prove to be very useful and complementary tools in the measurement of diffraction in $pp$ collisions.

The diffractive production of vector mesons and single hard photons at HERA provides a way to select and determine different generalized parton distributions (GPDs), the quantum mechanical parton wavefunctions, of the proton (S.~Wallon, P.~Kroll, M.~Diehl). Early results from Hermes (C.~Hadjidakis) and from H1 and Zeus (A.~Bruni, L.~Favart) seem to be in good agreement with current GPD models. M.~Diehl used this information to analyze the helicity structure of the pomeron, and A.~Donnachie discussed evidence for helicity-flip amplitudes at high energies. A number of talks discussed the physics of other exclusive diffractive reactions such as two-photon collisions which are sensitive to the vector meson distribution amplitudes as well as the exchange mechanism. Double-charm production, an indication of an intrinsic heavy charm component in the proton wavefunction, was demonstrated by the SELEX experiment at Fermilab (J.~Russ). The implications of intrinsic charm and bottom for hadron phenomenology and even Higgs production at large longitudinal momentum fractions was discussed by J.~Soffer. The production of pentaquarks and other exotic quark bound states was reviewed (S.~Ricciardi of BaBar, P.~Aslanyan) with the conclusion that the situation is still very confused with seemingly contradictory results, possibly because their production mechanism may strongly depend on the particulars of the initial state.

In view of the special occasion, a number of overviews were presented surveying the progress in diffractive physics, both experimentally and theoretically, made over the past twenty years. A.D.~Krisch told the story of his pioneering measurements of hadronic spin effects, and the extraordinarily large spin correlations that were discovered in large angle elastic $pp$ scattering and are still only partially understood. K.~Goulianos gave an overview of diffraction, both soft and hard, as measured at the Tevatron and its connection to HERA results. G.~Ingelman presented the enormous evolution in understanding of hard diffraction since hard diffraction was first observed by UA8 at the CERN $p\bar{p}$ collider.

The overall atmosphere of the conference was one of expectation: for a substantial increase of data from the new runs that have started at the Tevatron and at HERA~II, and from the LHC which will start running around the time of the next Blois Conference. Indeed, the next two Blois conferences may well see significant progress in the fundamental understanding of soft and hard diffraction.

\section{Acknowledgements}
The authors thank the conference organizers for their hospitality and the perfect organization of this anniversary conference in a beautiful venue. They thank all the conference speakers for their contributions to this conference summary.  References to the original literature may be found in the contributions of the speakers to these proceedings.

\end{document}